\documentclass[11pt]{article}
\usepackage{fullpage}
\usepackage{hyperref}
\usepackage{amsthm}
\newenvironment{IEEEproof}{\begin{proof}}{\end{proof}}

\usepackage[T1]{fontenc}
\usepackage[utf8]{inputenc}
\usepackage[british]{babel}
\usepackage[numbers]{natbib}
\usepackage{amsmath}
\usepackage{amsfonts} 
\usepackage{MnSymbol} 

\usepackage{footnote}

\usepackage[font=footnotesize]{subfig}
\usepackage{pgfplots}
\usepgfplotslibrary{units,groupplots}
\usetikzlibrary{patterns}
\pgfplotsset{compat=newest}

\usepackage{xspace}
\usepackage{multirow}
\usepackage{booktabs} 

\newtheorem{thm}{Theorem}
\newtheorem{lem}[thm]{Lemma}
\newtheorem{cor}[thm]{Corollary}

\usepackage{color}
\newcommand{\seq}[1]{\left\langle #1\right\rangle}

\newcommand{\sceil}[1]{\lceil #1\rceil}
\newcommand{\ceil}[1]{\left\lceil #1\right\rceil}

\newcommand{\set}[1]{\left\{ #1\right\}}

\newcommand{\Def}{:=}

\newcommand{\card}[1]{\left\vert{#1}\right\vert}

\newcommand{\nat}{\mathbb{N}}

\newcommand{\real}{\mathbb{R}}

\newcommand{\integer}{\mathbb{Z}}

\newcommand{\realrange}[2]{\left[#1, #2\right]}

\newcommand{\unitrange}[2]{\realrange{0}{1}}

\newcommand{\prob}[1]{{\mathbf{P}}\left[#1\right]}

\newcommand{\Oh}[1]{\mathcal{O}\!\left( #1\right)}

\newcommand{\Ohsmall}[1]{\mathcal{O}(#1)}
\newcommand{\oh}[1]{\mathrm{o}\!\left( #1\right)}

\newcommand{\llabel}[1]{\label{\labelprefix:#1}}
\newcommand{\labelprefix}{} 

\newcommand{\discussionsize}{\small}

\newenvironment{code}{\noindent\normalsize
\begin{tabbing}%
\hspace{2em}\=\hspace{2em}\=\hspace{2em}\=\hspace{2em}\=\hspace{2em}\=%
\hspace{2em}\=\hspace{2em}\=\hspace{2em}\=\hspace{2em}\=\hspace{2em}\=%
\kill}{\end{tabbing}}

\newcommand{\labelcommand}{}
\newcommand{\captiontext}{}
\newsavebox{\codeparam}
\newcounter{lineNumber}
\newenvironment{disscodepos}[3]{%
\renewcommand{\labelcommand}{#2}%
\renewcommand{\captiontext}{#3}%
\sbox{\codeparam}{\parbox{\textwidth}{#3}}%
\begin{figure}[#1]\begin{center}\begin{code}\setcounter{lineNumber}{1}}{%
\end{code}\end{center}\caption{\llabel{\labelcommand}\captiontext}\end{figure}}

{\end{disscodepos}}

\newcommand{\Declare}[2]{#1\mbox{ \rm : }#2}

\newcommand{\BR}[1]{\texttt{[}#1\texttt{]}}

\newcommand{\Function} {{\bf def\ }}
\newcommand{\Funct}[3]{\Function #1\Declare{{\rm (}{#2\rm )}}{#3}}

\newcommand{\Foreach}      {{\bf foreach\ }}
\newcommand{\Is}{\mbox{\rm := }}

\newcommand{\If}       {{\bf if\ }}

\newcommand{\Then}     {{\bf then\ }}

\newcommand{\Return}   {{\bf return\ }}

\newcommand{\RRem}[1]   {\`{\bf --\hspace{0.5mm}--~}{\rm#1}}

\newdimen\endofsize\endofsize=0.5em

\newcommand{\Tstart}{\alpha}
\newcommand{\Tbit}{\beta}
\newcommand{\Tword}{\Tbit} 

\newcommand{\eg}{\emph{e.g.}\xspace}
\newcommand{\ie}{\emph{i.e.}\xspace}
\newcommand{\keyval}{\ensuremath{(\textit{key},\textit{value})}\xspace}
\newcommand{\keyvalcount}{\ensuremath{(\textit{key},\textit{value},\textit{count})}\xspace}
\newcommand{\T}[1]{T_{\textit{#1}}}
\newcommand{\TI}{\!\times\!}
\newcommand{\op}[1]{\ensuremath{\mathrm{#1}}}

\makeatletter
\def\NAT@spacechar{~}
\makeatother

\makeatletter
\pgfdeclarepatternformonly[\GridSize]{MyFill}{\pgfqpoint{-1pt}{-1pt}}{\pgfqpoint{\GridSize}{\GridSize}}{\pgfqpoint{\GridSize}{\GridSize}}%
{
  \pgfsetcolor{\tikz@pattern@color}
  \pgfsetlinewidth{0.3pt}
  \pgfpathmoveto{\pgfqpoint{0pt}{0pt}}
  \pgfpathlineto{\pgfqpoint{\GridSize}{\GridSize}}
  \pgfusepath{stroke}
}
\makeatother
\newdimen\GridSize
\tikzset{
    GridSize/.code={\GridSize=#1},
    GridSize=4.25pt
}

\makeatletter
\pgfplotsset{
    groupplot xlabel/.initial={},
    every groupplot x label/.style={
        at={($({\pgfplots@group@name\space c1r\pgfplots@group@rows.west}|-{\pgfplots@group@name\space c1r\pgfplots@group@rows.outer south})!0.45!({\pgfplots@group@name\space c\pgfplots@group@columns r\pgfplots@group@rows.east}|-{\pgfplots@group@name\space c\pgfplots@group@columns r\pgfplots@group@rows.outer south})$)},
        anchor=north,
    },
    groupplot ylabel/.initial={},
    every groupplot y label/.style={
            rotate=90,
        at={($({\pgfplots@group@name\space c1r1.north}-|{\pgfplots@group@name\space c1r1.outer
west})!0.5!({\pgfplots@group@name\space c1r\pgfplots@group@rows.south}-|{\pgfplots@group@name\space c1r\pgfplots@group@rows.outer west})$)},
        anchor=south
    },
    execute at end groupplot/.code={%
      \node [/pgfplots/every groupplot x label]
{\pgfkeysvalueof{/pgfplots/groupplot xlabel}};
      \node [/pgfplots/every groupplot y label]
{\pgfkeysvalueof{/pgfplots/groupplot ylabel}};
    }
}
\def\endpgfplots@environment@groupplot{%
    \endpgfplots@environment@opt%
    \pgfkeys{/pgfplots/execute at end groupplot}%
    \endgroup%
}
\makeatother

\begin{document}

\title{Communication Efficient Checking of\\Big Data Operations}

\author{Lorenz Hübschle-Schneider and Peter Sanders\\
  Institute of Theoretical Informatics\\
  Karlsruhe Institute of Technology, Germany\\
  \texttt{\{huebschle,sanders\}@kit.edu}}

\maketitle

\begin{abstract}
  We propose fast probabilistic algorithms with low (\ie, sublinear in the
  input size) communication volume to check the correctness of operations in Big
  Data processing frameworks and distributed databases.  Our checkers cover many
  of the commonly used operations, including sum, average, median, and minimum
  aggregation, as well as sorting, union, merge, and zip.  An experimental
  evaluation of our implementation in Thrill (Bingmann et al., 2016) confirms
  the low overhead and high failure detection rate predicted by theoretical
  analysis.
\end{abstract}


\section{Introduction}

Recently, Big Data processing frameworks like Apache Spark~\cite{Spark}, Apache
Flink~\cite{Flink} and Thrill~\cite{Thrill} have surged in popularity and are
widely used to process of large amounts of data.  Computation and data are
distributed over a large number of machines connected by a fast network, and
processing is based on collective operations that transform datasets.  As
compute performance and memory capacity continue to grow (``Moore's Law'') while
simultaneously becoming ever cheaper, oftentimes all data can be held in main
memory.  Combined, these advances let users process enormous amounts of data
quickly.  However, as the number of machines increases, so does the rate of
hardware failures and thus the importance of \emph{fault tolerance}, as well as
the difficulty of writing correct programs that handle all edge cases.  Some
frameworks, \eg Apache Spark, can deal with failing machines~\cite{Spark}, but
none of the popular solutions can detect \emph{silent errors}.  These can be the
result of subtle errors in the programming, but also spontaneous bitflips in
memory (``soft errors''), caused for example by cosmic
rays~\cite{CosmicRays}---a concern that can largely be mitigated by using ECC RAM, but
at the cost of increased hardware expenditure.  We propose probabilistic algorithms
to detect silent failures in the frameworks' operations with
little overhead, which we refer to as \emph{checkers}.  These checkers verify the
integrity of the computation while treating the operation as a black box, \ie
independent of its implementation.  Our checkers cover many of the commonly used
operations of current-generation data-parallel computing frameworks.

Consider the options available to a programmer who wants to increase users'
confidence in the correctness of her program.  Formal verification would be
ideal, but very difficult for complex programs and does not address hardware errors.  Testing can help the
programmer to avoid mistakes, but cannot prove the absence of bugs.  Lastly, she
can write a small and fast program that verifies the output of the main program: a
\emph{checker}.  The checker must be very fast to avoid large slowdowns, much
faster than the main program.  The probabilistic checkers we consider here
make a favorable trade-off between confidence and speed.

In distributed computing, communication latency and bandwidth limitations are
among the main limiting factors.  Our main optimization criterion is therefore
the maximum amount of data sent or received at any single processing element
(PE), as the slowest PE determines overall running time.
In a previous paper~\cite{SSM13} we proposed to intensify the search for
algorithms with low \emph{bottleneck} communication volume.  More precisely,
consider an input of $n$ fixed-size elements distributed over the $p$~PEs such
that every PE holds $\Oh{n/p}$ elements.  Then we want \emph{no} PE to send or
receive more than $\oh{n/p}$ elements, or be part in more than a polylogarithmic
number of data exchanges (messages).

In this paper, we adopt the terminology of Thrill~\cite{Thrill} for operations.
We also design our checkers to become part of it, and implemented them within
Thrill for our experiments.

Note that the majority of our results also apply to distributed database
systems.  The proximity between data-parallel big data processing frameworks and
distributed databases is exemplified by the SQL layer of Apache
Spark~\cite{SparkSQL}.

Table~\ref{tbl:results} lists our main results.  For each operation, it states
whether the entire result needs to be available at all PEs (or a distributed
result suffices), whether the checker requires a certificate---and if yes,
whether a distributed certificate suffices---and the checker's running time.

\newcommand{\Problem}[2][2]{\multirow{2}{4.75cm}{#2}}
\newcommand{\ML}[2][2]{\multirow{#1}{*}{#2}}
\newcommand{\Row}[5][2]{\Problem[#1]{#2} & \ML[#1]{#3} & \ML[#1]{#4} & \ML[#1]{#5} \\\\}
\begin{table}
  \caption[Our main results]{Our main results. Parameters: input size $n$;
    number of processing elements (PEs)~$p$; failure probability~$\delta$;
    machine word size~$w$; trade-off parameter~$d$; communication startup
    cost~$\Tstart$; communication cost per bit~$\Tbit$}
  \label{tbl:results}

\centering
\begin{tabular}{lccl}\toprule
  \ML{\textbf{Operation}} & \multirow{2}{2cm}{\centering \textbf{Broadcast Result?}} & \multirow{2}{2cm}{\centering \textbf{Certificate  Required?}} & \ML{\textbf{Checker Running Time $\Oh{\cdot}$}}\\\\\midrule

  \Row{Sum/Count Agg.}{no}{no}
  {$\left(\frac np + \Tbit d w\right)\log_d\frac1\delta + \Tstart\log p$}

  \Row{Average Agg.}{no}{dist}%
  {same as above}

  \Row{Median Agg.}{yes}{yes$^{\mathrm{a}}$}%
  {same as above}

  \Row{Minimum Agg.}{yes}{yes}
  {$\frac np + \Tstart \log p$}

  \Row{Permutation, Sort, Union, Merge, Zip, GroupBy$^{\mathrm{b}}$, Join$^{\mathrm{b}}$}{no}{no}
  {$\left(\frac{n}{pw} + \Tbit\right) \log\frac1\delta + \Tstart \log p$}\\
  \bottomrule\\[-0.8em]
  \multicolumn{4}{l}{$^{\mathrm{a}}$ no certificate required if input elements are distinct.}\\
  \multicolumn{4}{l}{$^{\mathrm{b}}$ invasive checker for input redistribution phase.}
\end{tabular}
\end{table}

\section{Preliminaries}\label{s:prelim}

Let an operation's input data set consist of $n$ elements, each represented by a
fixed number of machine words\footnote{Oftentimes, adaptation to variable sized
  objects is possible.  Our focus on fixed size objects is mainly to simplify
  notation.}, and let $k$ denote the number of elements in the output of a
particular operation, and let $w$ be the machine word size in bits.  Consider
$p$ processing elements (PEs) connected by a network, numbered $1..p$, where
$a..b$ is shorthand for $\{a,\ldots,b\}$ throughout this paper.  We assume that
PEs can send and receive at most one message simultaneously (full-duplex,
single-ported communication).  Then, sending a message of size $m$ bits takes
time $\Tstart + \Tbit m$, where $\Tstart$ is the time to initiate a connection
and $\Tbit$ the time to send a single bit over an already-established
connection.  We treat $\Tstart$ and $\Tbit$ as variables in asymptotic
analysis.  Thus, a running time of $\Oh{x + \Tbit y + \Tstart z}$ allows us to
discuss \emph{internal work} $x$, \emph{communication volume} $y$, and the
\emph{number of messages} (or \emph{latency}) $z$ separately.  Frequently, all
three aspects are important, and combining them into a single expression for
running time allows for a concise representation.

Throughout this paper, logarithms with an unspecified base are binary
logarithms: $\log{x} \Def \log_2{x}$.

\paragraph{Error Model} Checkers have one-sided error: they are
never allowed to reject the result of a correct computation.  Only in the case
of an incorrect result are they allowed to erroneously accept with a small
probability, limited by the parameter $\delta>0$.

\paragraph{Distribution of Input} We usually assume that the $n$ input elements
are distributed over the $p$ PEs such that each PE holds $\Oh{n/p}$ elements.
This is mainly in order to simplify notation. A generalization would usually involve an additional term proportional to the maximum number of elements located on any PE.
Note that we explicitly \emph{do not} assume random distribution of
the input, as that would require redistribution of the entire input in cases
where this is not the case already.

\paragraph{Certificates} Some operations become much easier to check if the
result of the operation is accompanied by a \emph{certificate} to facilitate
checking. However, when an algorithm provides an output along with a certificate
for said output, the certificate might also be faulty.  We need to take care not
to accept incorrect results because the certificate contains the same flaw, \ie
we need to verify the correctness of the certificate as well.

\paragraph{Collective Communication} A \emph{broadcast} distributes a message
to all PEs.  \emph{Reduction} applies an associative operation to a sequence of
$k$ bits.  In an \emph{all-reduction}, the result is also broadcast.
All of these operations can be performed in time
$\T{coll}(k) \Def \Oh{\Tbit k + \Tstart \log p}$~%
\cite{BalEtAl95,SST09}.  In an \emph{all-to-all} communication, every PE
sends $k$ bits to every other PE.  This can be performed in
$\T{all-to-all}(k) \Def \Oh{\Tbit k + \Tstart p}$ using direct
delivery or $\Oh{\Tbit k \log p + \Tstart \log p}$ using hypercube indirect
delivery~\cite{MehSanPar}

\paragraph{Hashing} To simplify analysis, we assume the availability of
\emph{random hash functions}, \ie hash functions chosen uniformly at random
from the set of all mapping between the input and output value types.  Then,
hash values can be treated like numbers chosen uniformly at random from the hash
functions' image space.  We explain separately when weaker guarantees on the
randomness of the hash function suffice.

\paragraph{Reduction} A \emph{reduce} operation collects elements by their
keys, processing all elements of the same key in an arbitrary order using an
associative reduce function
$f: \textit{Value} \times \textit{Value} \rightarrow \textit{Value}$.  This
function describes how two combine two elements into one.  To reduce the local
elements of a PE, we use a hash table~$h$.  Process elements one-by-one,
denoting the current element by $e=(k,v)$, and set $h[k] = f(v, h[k])$ or
$h[k] = f(v, 0)$ if $h[k]$ does not exist.  We then use a simple reduction
algorithm (see ``Collective Communication'' above) to obtain the final result at
PE~0.  The total time taken is
$\Oh{n/p + \T{coll}(wk)} = \Oh{n/p + \Tbit wk + \Tstart \log p}$.

\paragraph{GroupBy} A more general approach to aggregation is \emph{GroupBy},
where all elements with a certain key are collected at one PE and processed by
the group function $g: [\textit{Value}] \rightarrow \textit{Value}$.  This
enables the use of more powerful operators such as computing median, but
requires more communication.  Total running time is
$\Oh{n/p + \T{all-to-all}(wn/p)} = \Oh{n/p + \Tbit wn + \Tstart p}$.

\paragraph{Result Integrity} When the output of an operation or
a certificate is provided at all PEs rather than in distributed form, we need to
ensure that all PEs received the \emph{same} output or certificate.
This can be achieved by hashing the data in question with a random hash function, and
comparing the hash values of all other PEs.  This can be achieved in time
$\Oh{k+\Tstart\log p}$ by broadcasting the hash of PE 0, which every PE can
compare to its own hash, and aborting if any PE reports a difference.

\section{Related Work}\label{s:related}

We were surprised to find that distributed, let alone communication efficient,
probabilistic checkers appear to be a largely unstudied problem.  We therefore
give a short overview of techniques for increasing confidence in the output of a
program.

From a theoretical standpoint, formally verifying the correctness of a program
is the ultimate goal.  While tremendous progress has been made in this area
(e.g.~\cite{KeYBook2016}), verifying implementations of complex distributed
algorithms in popular programming languages is still infeasible.

In practice, programs are often tested against a set of test cases of
known good input/output pairs, ensuring that the program computes the correct result
without crashing for all test cases. However, testing can never demonstrate the
\emph{absence} of bugs.

\citet{Blum89Checkers} introduce probabilistic checkers in a sequential setting.
\citet{MMNS11} build upon this to design (sequential) \emph{certifying algorithms}, stressing
the importance of simplicity of the checker, and putting increased focus on
certificates (also termed \emph{witnesses}) that prove that the output is
correct.  While formally verifying the algorithm may be infeasible, doing so for
the verifier may be within reach.  As a result, one can use fast but complicated
algorithms that may be beyond the programmer's competence to fully understand,
while maintaining certainty in the correctness of the result.

Verification of arbitrary computations performed by a single machine or
outsourced to a set of untrusted machines is a well-studied
problem, dating back over 25 years and published under names such as
``verifiable computing''~\cite{Gennaro10verifiable}, ``checking
computations''~\cite{babai91check}, or ``delegating
computations''~\cite{GKR08delegating}.  All of these are computationally
expensive beyond the limits of feasibility, despite recent efforts to make
verifiable computing more practical~\cite{PHGR13pinocchio}.  A recent survey
by \citet{WB15Verifying} concludes that \emph{``[t]he sobering news,
  of course, is these systems are basically toys''} due to orders-of-magnitude
overhead.  In contrast, our work focuses on checking specific operations,
allowing us to develop checkers that are fast in practice.

An approach that has received significant attention for linear algebra
kernels is \emph{Algorithm-Based Fault Tolerance}
(ABFT)~\cite{HuangAbr84,ABFThpc}.  By encoding the input using checksums, and
modifying the algorithms to work on encoded data, ABFT techniques detect and
correct any failure on a single processor.  However, this requires the
algorithms to be redesigned to operate on the encoded data.

Note that while some existing systems, \eg Apache Spark~\cite{Spark},
implement some fault tolerance measures such as detecting and handling the
failure of individual nodes, there are large classes of failures that
no existing big data processing system appears to detect or mitigate.
These include data corruption caused by \emph{soft
  errors}, \ie incorrectly handled edge cases in the programming, or bitflips in
CPU or main memory caused by cosmic rays, leak voltage, as well as \emph{hard errors},
\ie device defects causing repeated failure at the same memory locations.  A
common solution to memory errors is the use of ECC RAM~\cite{GoogleDC}, which
can correct most soft errors and detect (but often not correct) many hard
errors~\cite{CosmicTwice}.  This is orthogonal to our checkers: a checker can
never detect errors introduced into the input data before the algorithm is
invoked, so the use of ECC RAM remains advisable even when computations can be
checked.

\section{Count and Sum Aggregation}\label{s:sumagg}

\newcommand{\imin}{\ensuremath{\breve{\imath}}}
\newcommand{\rmin}{\ensuremath{\hat{r}}}
\newcommand{\hb}[1]{\ensuremath{\overline{h_{#1}}}}
\newcommand{\hbb}[1]{\ensuremath{\overline{h'_{#1}}}}
\newcommand{\Fb}{\ensuremath{\overline{F}}}

Reductions are perhaps the most important operation in the kind of Big Data
systems we consider, and the paradigm they extend even carried them in their name:
MapReduce.  The checker we describe works not only for sum aggregation, but also
other operations on integers that fulfill certain properties.  We require that
the reduce operator $\oplus$ be associative, commutative, and satisfy
$x \oplus y \neq x$ for all $y\neq 0$, \ie every element except the neutral
element changes the result.  Examples include count aggregation, which
conceptually equals sum aggregation where the value of every element is mapped
to 1, and exclusive or (\emph{xor}).  Without loss of generality, we therefore
only discuss sum aggregation in this section.  Several other common choices of
reduce functions violating the above requirements and aggregations requiring the
additional power of GroupBy are discussed in Section~\ref{s:more}.

In sum aggregation, we are given a distributed set of \keyval-pairs.  Let $K$ be
the (unknown) set of keys in the input and $k \Def \card{K}$ its equally unknown
size.  The output of sum aggregation then consists of a single value for each
key, which is the sum of all values associated with it in the input.  In SQL,
this operation would be expressed as
\[\texttt{SELECT key, SUM(value) FROM table GROUP BY key}.\]

\begin{thm}\label{thm:sumagg}
  Let $\oplus$ be an associative and commutative reduce function on integers
  satisfying $x \oplus y \neq x$ for all elements $x,y$ with $y \neq 0$.  Then
  $\oplus$-aggregation of $n$ elements with failure probability at most $\delta$
  can be checked in time
  $
  \T{check-sum}(n, p, \delta) \Def \Oh{\left(\frac np + \Tbit d w\right)
    \log_d\frac1\delta + \Tstart\log p}$, where $d$ is a tuning parameter.
\end{thm}

Note that only the local work depends on the input size, and that the running
time is independent of the number of keys.

To check the result of such an aggregation, we apply a na\"ive sum reduction
algorithm to a condensed version of the input.  We shall sometimes refer to this
as the \emph{minireduction} in subsequent sections.  Let $d$ from the statement
of Theorem~\ref{thm:sumagg} be the size of the condensed key space, with
$2\leq d\ll k$.  We then use a random hash function $h: K \rightarrow 1..d$ to
map the original keys to the reduced key space.  Let $\rmin\geq d$ be a modulus
parameter, with $r$ chosen uniformly at random from the half-open interval
$(\rmin,2\rmin]$.  Apply a na\"ive sum reduction modulo~$r$ to the thus-remapped
input and---separately---the output of the aggregation algorithm.  If both
produce the same result, the operation was likely conducted correctly.
Pseudocode is given in Algorithm~\ref{alg:sumagg}.

\renewcommand{\figurename}{Algorithm}
\begin{figure}
\begin{code}
  \Funct{checkSumAgg}{%
    \Declare{$v$}{Element\BR{$n_i$}},
    \Declare{$o$}{Element\BR{$k_i$}},
    \Declare{$\delta$}{$\real$}}{Boolean}\+\\
  $\!(d, \rmin):\nat\times\nat\ \Is$numerically determined parameters (see Table~\ref{tbl:num_min})\\
  $r: \nat\ \Is$random number from \rmin+1..2\rmin\RRem{modulus parameter}\\
  $h: K \rightarrow 1..d\ \Is$ random hash function\RRem{maps keys to buckets} \\
  $w_v\ \Is \text{cRed}(v, d, r, h)$\RRem{apply condensed reduction to input}\\
  $w_o\ \Is \text{cRed}(o, d, r, h)$\RRem{...and asserted result of sum aggregation}\\
  \Return $w_v=w_o$\-\RRem{significant only at PE 0}\\\\

  \Funct{cRed}{%
    \Declare{$\textit{arr}$}{Element\BR{}},
    \Declare{$d$}{$\nat$},
    \Declare{$r$}{$\nat$};
    \Declare{$h$}{$K \rightarrow 1..d$}}{Value\BR{$d$\kern 0.08em}}\+\\
  $t: \text{Value}\texttt{[$d$]} = \seq{0,\ldots,0}$ \\
  \Foreach{$(k,v) \in$ \textit{arr}}\+\\
    $t[h[k]]\ \Is (v + t[h[k]]) \text{ mod } r$\-\RRem{local reduction}\\
  \text{Reduce}($t$, $+_{r}$, 0) \RRem{reduce to PE 0 with addition modulo $r$}\\
  \Return $t$ \RRem{significant only at PE 0}
\end{code}
\caption{\label{alg:sumagg} A single iteration of the sum aggregation checker.
  Here, $v$ is the input to the sum aggregation operation, a sequence of $n$
  elements of which PE $i$ holds $n_i$; $o$ is the asserted result with $k_i$
  out of $k$ elements at PE $i$; and $\delta$ is the maximum allowed failure
  rate.}
\end{figure}

\begin{lem}\label{lem:sumagg_prob}
  A single iteration of the above sum aggregation checker fails with probability
  at most~$\frac1\rmin + \frac1d$.
\end{lem}
\begin{IEEEproof}
  If the aggregation was performed correctly, then the per-bucket results in the
  reduced keyspace are the same for input and output, and thus the checker
  always accepts a correct result.  Thus assume from now on that the output of
  the aggregation operation is \emph{incorrect}, \ie the checker \emph{should}
  fail.

  For $i \in K$, let $n_i$ be the (correct, unknown) $\oplus$-aggregate of all values with
  key $i$, and $n_i'$ be the asserted such value.  Then, as the result is
  incorrect, there exists at least one $i$ with $n_i \neq n_i'$.  Let
  $I \Def \{i \in K \mid n_i \neq n_i'\}$ be the keys whose results were
  computed incorrectly by the operation.

  Let $h: K \rightarrow 1..d$ be the random hash function used to map keys to
  the condensed keyspace. We use this hash function to map elements to the $d$
  buckets by their keys, and reduce the values in associated counters modulo
  $r$, where $r$ is chosen uniformly at random from $(\rmin,2\rmin]$.
  Effectively, we operate in the residue class ring $\integer/r\integer$.  Thus
  the checker fails if
  \[ \forall_{j \in 1..d}:\bigoplus_{\substack{i\in I\\h(i)=j}}{n_i} =
    \bigoplus_{\substack{i \in I\\h(i)=j}}{n_i'} \mod r. \]

  We prove the claimed failure probability by first considering the failure
  modes introduced by the modulus, and then analyzing a version of the
  checker without a modulus.  In this second part, we injectively map each hash
  function $h$ for which the checker fails to $d-1$ distinct hash functions for
  which it does not fail.

  \textbf{(1)} If there is an $i\in I$ with $n_i = n_i' \mod r$, the checker
  cannot detect the error in this key.  For a single key, this occurs with
  probability $r^{-1}\leq\rmin^{-1}$ by definition of~$r$.  However, for the
  checker to fail, $n_i=n_i'\mod r$ must hold for all $i\in I$. Thus, the
  probability of failure declines exponentially with~$\card{I}$, and
  $\card{I}=1$ is the hardest case.  Therefore, $\rmin^{-1}$ bounds the total
  additional failure probability introduced by the modulus.

  \textbf{(2)} Define
  $F := \set{h: K \rightarrow 1..d \mid \text{checker fails for\ } h}$ and let
  $\imin := \min I$ be the first key with mismatched aggregate value.  For each
  $h \in F$, we define the $d-1$ hash functions $\hb{j}$ with
  \[ \forall_{j \in 1..d \setminus \{h(\imin)\},\ i \in K}: \hb{j}(i) =
    \begin{cases}
      h(i) & i \neq \imin,\\
      j & \textit{else}
    \end{cases}
  \]
  and let
  $\Fb := \set{ \hb{j} \mid h \in F, j \in 1..d \setminus \set{h(\imin)} }$.
  Clearly, the checker does not fail for any $\hb{j} \in \Fb$, as exactly one
  element with different values is being remapped in a hash function for which
  it does fail (the case of $n_{\imin}=n'_{\imin}\mod r$ is treated
  in~\textbf{(1)}).  By the assumption $x\oplus y \neq x$ for $y \neq 0$, the
  result will differ in exactly two buckets, and the checker will notice.

  We also need to show that the mapping is injective, \ie that the new hash
  functions are unique and thus $|\Fb| = (d-1) \card{F}$.  Clearly, for given
  $h$ all of its $\hb{j}$ are different, so assume that there exists an
  $h' \neq h$ and $j, j'$ so that $\hb{j} = \hbb{j'} \in \Fb$.  Then, by
  definition, $j = \hb{j}(\imin) = \hbb{j'}(\imin) = j'$ and thus $j=j'$.  For
  $\hb{j} = \hbb{j}$, we furthermore need
  \[ \forall_{i \in K}: \hb{j}(i) = \hbb{j}(i) \
    \stackrel{\textit{def.}}{\Longleftrightarrow}\ %
    \forall_{i \in K\setminus\{\imin\}}: h(i) = h'(i)
  \]
  Thus $h=h'$ if and only if $h(\imin) = h'(\imin)$.  But this must hold, for
  otherwise we would have $h' = \hb{h'(\imin)} \in \Fb$ by definition of the
  $\hb{j}$, which contradicts the assumption that $h' \in F$.  Therefore, such
  an $h'$ cannot exist and $|\Fb| = (d-1) \cdot \card{F}$.
\end{IEEEproof}

The failure probability bound in \textbf{(2)} is tight: if the only difference
between the two inputs is the key of a single item, then the probability that
the hash values of the new and old key are the same (and the modification thus
goes unnoticed) is $1/d$.  This follows from the uniformity of random hash
functions.

\begin{lem}\label{lem:sumagg_time}
  Checking $\oplus$-reduction with $\oplus$ as in Theorem~\ref{thm:sumagg},
  $n$~input pairs, and $k$ keys, using $d$ buckets, with moduli in $(\rmin,2\rmin]$
  and probability of failure at most $\delta>0$, is possible in time
  \[
    \Oh{\left(\frac{n}{p} +\Tbit d \log
      (2\rmin)\right) \log_{\left(\frac1\rmin+\frac1d\right)^{-1}}{\delta^{-1}} + \Tstart\log p}.
  \]
\end{lem}
\begin{IEEEproof}
  From the above description we can see that a single iteration of the checker
  requires time $\Oh{n/p}$ to hash and locally reduce the input, and
  $\T{coll}(d\log(2\rmin)) = \Oh{\Tword d\log(2\rmin) + \Tstart\log p}$ for the
  reduction.  By repeating the procedure
  $\lceil\log_{\left(\frac 1\rmin + \frac1d\right)^{-1}}{\delta^{-1}}\rceil$
  times, we can increase the probability of detecting an incorrect result to at
  least $1-\delta$ by accepting only if all repetitions of the procedure declare
  the result to be likely correct.  To keep the number of messages to a minimum,
  we can execute all instances of the checker simultaneously (this also means
  that we only have to read the input once), and perform their reductions at the
  same time.
\end{IEEEproof}

We can instantiate this checker in different ways to obtain the characteristics
we wish.  First, we use this to show Theorem~\ref{thm:sumagg}:

\begin{IEEEproof}[Proof (Theorem~\ref{thm:sumagg})]
  The theorem follows from Lemma~\ref{lem:sumagg_time} with
  $d \leq \rmin\leq 2^{w-1}$ and $(\frac1\rmin+\frac1d)^{-1} < d$.
\end{IEEEproof}

We can also minimize bottleneck communication volume and find that minimum at
$d=2$ buckets, $\rmin=8$ for a modulus range of $9..16$, and thus a
minireduction result size of 8 bits with $\log_{1.6}{\delta^{-1}}$ repetitions.
However, this high number of repetitions causes a lot of local work.  Further,
the practical usefulness of sending single bytes across the network is
questionable.  In effect, real-world interconnects have an effective minimum
message size $b$, such that sending fewer than $b$ bits is not measurably faster
than sending a message of size $b$ bits.  Thus, our goal should be to minimize
the number of iterations---%
$\ceil{\log_{\frac1\rmin+\frac1d}{\delta}}$---under the constraint that the
result size be close to $b$ bits:
$d\ceil{\log (2\rmin)} \sceil{\log_{\frac1\rmin+\frac1d}{\delta}} \leq b$.  This
relation can be used to (numerically) compute optimal choices of $\rmin$ and $d$
for given~$b$.  Table~\ref{tbl:num_min} shows such values for some interesting
choices of $b$ and~$\delta$.  However, in practice, keeping local work low might
be more important than these solutions to minimize $\delta$ admit, and one might
prefer to trade a reduced number of iterations for a larger value of $d$ and
perhaps choose $\rmin=2^{31}$.

\begin{table}[bt]
  \captionsetup{justification=centering}
  \caption{Some numerically determined optimal values for bucket\\count $d$ and modulus parameter $\rmin$ given a message size of $b$ bits. \label{tbl:num_min}}
  \centering
  \begin{tabular}{rrrrrr}\toprule
    $b$ & $\delta$ & \quad\qquad $d$ & $\rmin$ & \#its & achieved $\delta$ \\\midrule
    $1024$ & $10^{-4}$ & 37 & $2^8$ & 3 & $3.0\cdot 10^{-5}$ \\
    $1024$ & $10^{-6}$ & 25 & $2^7$ & 5 & $2.5\cdot 10^{-7}$ \\
    $1024$ & $10^{-8}$ & 18 & $2^7$ & 7 & $4.1\cdot 10^{-9}$ \\
    $1024$ & $10^{-10}$ & 14 & $2^6$ & 10 & $2.5\cdot 10^{-11}$ \\
    $1024$ & $10^{-20}$ & 6 & $2^4$ & 32 & $3.3\cdot 10^{-21}$ \\\midrule
    $4096$ & $10^{-6}$ & 124 & $2^{10}$ & 3 & $7.4\cdot 10^{-7}$ \\
    $4096$ & $10^{-10}$ & 68 & $2^9$ & 6 & $2.1\cdot 10^{-11}$ \\
    $4096$ & $10^{-20}$ & 32 & $2^8$ & 14 & $4.4\cdot 10^{-21}$ \\\midrule
    $16\,384$ & $10^{-7}$ & 420 & $2^{12}$ & 3 & $1.8\cdot 10^{-8}$ \\
    $16\,384$ & $10^{-10}$ & 273 & $2^{11}$ & 5 & $1.2\cdot 10^{-12}$ \\
    $16\,384$ & $10^{-20}$ & 148 & $2^{10}$ & 10 & $7.6\cdot 10^{-22}$ \\
    $16\,384$ & $10^{-30}$ & 93 & $2^{10}$ & 16 & $1.3\cdot 10^{-31}$ \\\midrule
    $65\,536$ & $10^{-10}$ & 1170 & $2^{13}$ & 4 & $9.1\cdot 10^{-13}$ \\
    $65\,536$ & $10^{-20}$ & 630 & $2^{12}$ & 8 & $1.3\cdot 10^{-22}$ \\
    $65\,536$ & $10^{-30}$ & 420 & $2^{12}$ & 12 & $1.1\cdot 10^{-31}$ \\
    $65\,536$ & $10^{-40}$ & 321 & $2^{11}$ & 17 & $2.9\cdot 10^{-42}$ \\\bottomrule
  \end{tabular}
\end{table}

\paragraph*{Optimizations}  Multiple instances of this algorithm can be
executed concurrently by using a hash function that computes $c\cdot\ceil{\log d}$ bits.
Its value can then be interpreted as $c$ concatenated hash values for separate
instances, enabling bit-parallel implementation.  It is also possible to use
Single Instruction Multiple Data (SIMD) techniques to further reduce local work.
Refer to Section~\ref{s:exp} for specific implementation details.

\section{Permutation and Sorting}\label{s:sort}

Many approaches for permutation checking exist in the sequential case, and often
directly imply communication efficient equivalents.  Perhaps most elegantly and
first described by \citet{CarWeg81hash}---albeit in a less general manner than
we prove here---we can use a random hash function and compare the sum of hash
values in the input and output sequences.  Once we have established that a
sequence is a permutation of another, verifying sortedness of the output
sequence only requires that each PE receive the smallest element of its
successor PE and compare its local maximum to it.

\begin{lem}\label{lem:perm_hash}
  Let $E=\seq{e_1,\ldots,e_n}$ and $O=\seq{o_1,\ldots,o_n}$ be two sequences of
  $n$ elements from a universe $U$.  Then, for a random hash function
  $h: U \rightarrow 0..H-1$, define
  $\lambda:=\nobreak\sum_{i=1}^{n}{h(e_i)-h(o_i)}$ in
  $\mathbb{F}_H = \integer/H\integer$ (\ie, $\mathrm{mod}\ H$), and
  \[\prob{\lambda=0} = \begin{cases}1&\text{if $E$ is a permutation of $O$}\\
      H^{-1}& \text{otherwise.}\end{cases}\]
\end{lem}
\begin{IEEEproof}
  First define $h(X) \Def \sum_{x\in X}h(x)$ for any set or sequence $X$.  If
  $E$ and $O$ are permutations of each other, then $h(E) = h(O)$ due to
  commutativity of addition, and thus $\lambda\nobreak=\nobreak0$ by
  associativity.  Otherwise, when interpreted as sets, $E$ and $O$ each consist
  of a set of shared and private elements: $E=S\cup P_E$, $O=S\cup P_O$, with
  $P_E\cap P_O=\emptyset$.  By commutativity, the shared elements contribute the
  same value $h(S)$ to both sums of hash values: $h(E) = h(S) + h(P_E)$ and
  $h(O) = h(S) + h(P_O)$.  Since $P_E$ and $P_O$ are disjoint sets and $h$ is a
  random hash function, $h(P_E)$ and $h(P_O)$ are independent and uniformly
  distributed over $0..H-1$, and thus equal with probability $1/H$.  The same
  thus also holds for $h(E)$ and $h(O)$.
\end{IEEEproof}

Every PE can compute the sum for its $\Oh{n/p}$ local elements independently
(see Section~\ref{s:prelim}).  A sum all-reduction modulo~$H$ over these values
then yields $s$.  The algorithm can thus be executed in time
$\Ohsmall{n/p+\Tbit \log H + \Tstart\log p}$.

However, this algorithm requires confidence in the randomness of the hash
function $h$.
If we do not have access to a sufficiently trusted hash function, we can use a
different approach based on constructing a polynomial from $E$ and $O$, commonly
attributed to R.J. Lipton, and essentially a solution to Exercise 5.5 in
\citet{MehSan08}.

\begin{lem}\label{lem:perm_poly}
  For two sequences $E=\seq{e_1,\ldots,e_n}$ and $O=\seq{o_1,\ldots,o_n}$ from a
  universe $0..U-1$ and any $\delta>0$, choose a prime
  \( r > \max(n/\delta, U-1) \) and define the polynomial
  \[ q(z) := \prod_{i=1}^{n}{(z-e_i)} - \prod_{i=1}^{n}{(z-o_i)} \mod r.\]
  Then for random $z\in 0..r-1$, \[\prob{q(z)=0}\,
    \begin{cases}= 1&\text{if $E$ is a permutation of $O$}\\
      < \delta & \text{otherwise.}\end{cases}\]
\end{lem}
\begin{IEEEproof}
  If $E$ is a permutation of $O$, then both products have the same factors and
  thus $q(z)=0$ for all $z$ by commutativity of multiplication. Otherwise,
  because $r$ is larger than any $e_i$ or $o_i$ (this is important to ensure
  that no pair $i,j$ with $e_i \equiv o_j \mod r$ exists) is prime, $q$ is a
  non-zero polynomial of degree~$n$ in the field $\mathbb{F}_r$.  Such a
  polynomial has at most $n$ roots.\footnote{This can easily be seen by
    induction.  It is easy to verify for $n \leq 1$.  Now let~$q$ be a
    polynomial of degree $n\geq 2$, and $a$ be a root of $q$ (if none exists, we
    are finished).  Then $q=q'\cdot(X-a)$ for some polynomial $q'$ of degree
    $n-1$, and by the induction hypothesis, $q'$ has at most $n-1$ roots.  For
    some $b\in \mathbb{F}_r$, $q(b)=q'(b)\cdot(b-a)$ is zero iff $a=b$ or
    $q'(b)=0$.  Thus $q$ has at most $n$ roots.}
  Thus the probability of $q(z)=0$ is at most
  $\frac{n}{r}<\frac{n}{n/\delta}=\delta$.
\end{IEEEproof}


Instead of doing $2n$ expensive multiplication-modulo-prime operations, one
could also consider using carry-less multiplication in a
Galois Field~$\mathrm{GF}(2^\ell)$ with an irreducible polynomial.
Multiplication in Galois Fields can be implemented very efficiently, \eg using
Intel SIMD instructions~\cite{GFSIMD}.

\begin{thm}\label{thm:perm} It is possible to check whether a
  sequence of $n$ elements is a permutation of another such sequence with
  probability at least $1-\delta$ in
  time~$\Oh{(n/(pw) + \Tbit) \log\frac1\delta + \Tstart \log p}$
  $=: \T{check-perm}(n,p,\delta)$.
\end{thm}
\begin{IEEEproof}
  We can boost the success probabilities of the algorithms from
  Lemmata~\ref{lem:perm_hash} and~\ref{lem:perm_poly} arbitrarily by executing
  several independent instances and accepting only if all instances do.
  Batching communication keeps latency to $\log p$.  Choose $H=2^w$ in
  Lemma~\ref{lem:perm_hash} to obtain the claimed bound.  In
  Lemma~\ref{lem:perm_poly}, choose $\delta=2^{-w+1}n$.  Then, $r$ can always be
  chosen in $[2^{w-1},2^w]$ by Bertrand's postulate, and a machine word can hold
  the values.
\end{IEEEproof}

\begin{thm}\label{thm:sort}
  Checking whether a sequence of $n$ elements is a sorted version of another such
  sequence with probability $\geq 1-\delta$ is possible in time
  $\T{check-sort}(n,p,\delta) \Def
  \Oh{\T{check-perm}(n,p,\delta)}$.
\end{thm}
\begin{IEEEproof}
  After verifying the permutation property using Theorem~\ref{thm:perm}, it
  remains to be checked whether the elements are sorted.  First, verify that the
  local data is sorted in time~$\Oh{n/p}$.  Then, transmit the locally smallest
  element to the preceding PE, and receive the smallest element of the next PE.
  Compare this to the locally largest element.  Lastly, verify that no PE
  rejected using a gather operation.  In total, this requires time
  $\Oh{\T{check-perm}(n,p,\delta)} + \Oh{n/p} + 2\Tbit w + 2\Tstart +
  \T{coll} = \Oh{\T{check-perm}(n,p,\delta)}$.
\end{IEEEproof}

\section{Further Checkers}\label{s:more}

There are many operations for which we can construct checkers from the sum
aggregation and permutation checker.  We discuss several in the following
subsections.  Afterwards, we turn to some \emph{invasive} checkers, \ie checkers
that do not treat the operation as a black box, and instead check only part of
the operation.  While less desirable than true checkers, these checkers allows
us to broaden the range of covered operations.

\subsection{Average Aggregation}\label{ss:avg}

Computing the per-key averages is impossible to express in a scalar reduction,
but becomes easy when replacing \keyval-pairs with \keyvalcount-triples and
using the reduce function $\oplus$ with
$(k_1, v_1, c_1) \oplus (k_1, v_2, c_2) \Def (k_1, v_1+v_2, c_1+c_2)$.  The
computation is then followed up by an output function $h$ with
$h(k_1, v_1, c_1) \Def (k_1, v_1/c_1)$.  This avoids the use of the much more
communication-expensive GroupBy function.

Checking average aggregation is easy when these per-key element counts are
available in a certificate, for then we can reconstruct the result of a sum
aggregation by undoing the final division---termed $h$ above---by multiplying
the average value with the count for every key.  As described above, this
certificate naturally arises during computation anyway, and thus does not impose
overhead on the average computation.  Now we can leverage the sum checker,
applying it both to the input sequence and the reconstructed sums.  As the
product is computed component-wise, both the asserted averages and the
certificate can be supplied in distributed form, as long as both values are
available at the same PE for any key.

To prevent accidental mismatches when both averages and counts are scaled in a
way that yields the same reconstructed sums---\eg double the averages and halve
the counts---, we also need to check the correctness of the counts.  Therefore,
we also need to apply the \emph{count aggregation checker} to the input and the
certificate.  The same can also be achieved in a single step by applying the
\keyvalcount-triple aggregation trick using the reduce function $\oplus$
described above to the checker.

\begin{cor}\label{corr:avgagg}
  For a given input sequence of $n$ key-value pairs $\seq{e_1,\ldots,e_n}$ with
  $e_i=(k_i,v_i)$, keys $k_i \in K$, and $k\Def\card{K}$, it is possible to check whether a
  set of $k$ asserted per-key averages $\seq{a_1,\ldots,a_k}$ is correct if
  the number of values associated with each key is available as a certificate
  $\seq{c_1,\ldots,c_k}$, by supplying $\seq{(e_1,1),\ldots,(e_n,1)}$ as input
  and $\seq{(a_1\cdot c_1,c_1),\ldots,(a_k\cdot c_k,c_k)}$ as output to the sum
  aggregation checker of Section~\ref{s:sumagg}, achieving the time complexity and
  success probability of Theorem~\ref{thm:sumagg}.
\end{cor}

\subsection{Minimum and Maximum Aggregation}\label{ss:min}

Now consider computing the minimum or maximum value per key (\emph{w.l.o.g.},
we shall only consider minima henceforth).  This is a surprisingly
difficult operation to check.  Clearly, the sum aggregation checker of
Section~\ref{s:sumagg} is not directly applicable, as the $\min$ function does
not satisfy the requirement $\min(a,b)\neq a$ if $b\geq a$.  To check
$\min$-aggregation, we need to verify for each key that (a) no elements smaller
than the purported minimum exist, as well as determine whether (b) the minimum
value does indeed appear in the input sequence.  Both subproblems seem to
require the asserted result to be known at all PEs.  Let
$S=\seq{x_1,\ldots,x_n}$ be the input sequence, and $M=\seq{m_1, \ldots, m_k}$
the asserted output.  Given~$M$, property (a) is easy to verify by iterating the
locally present part of $S$ and verifying that no element exists with a value
smaller than the entry of its key in~$M$.

However, property (b) is surprisingly hard to check using $\oh{k}$ bits of
communication (it is easy to verify in time
$\Oh{n/p + \Tbit k + \Tstart \log p}$ using a bitwise $\mathrm{or}$
reduction---see Section~\ref{s:prelim}---on a bitvector of size $k$ specifying
which keys' minima are present locally, and testing whether each bit is set in
the result).  A certificate in the form of the locations of the minima,
available in full at every PE, remedies this.  Each PE then needs to verify its
set of local asserted minima.  The certificate is required to be
available in full at every PE so that we can ensure that all keys are
covered---otherwise, a faulty algorithm might simply ``forget'' a key, and the
checker would not be able to notice.

\begin{thm}\label{thm:minagg}
  Checking minimum (maximum) aggregation is possible in time
  $\Oh{n/p + \Tstart \log p}$ provided that the asserted output and a
  certificate specifying which PE holds the minimum (maximum) element for any
  key are available at all PEs.
\end{thm}
\begin{IEEEproof}
  Follows directly from the above discussion.
\end{IEEEproof}

Note that this checker is not probabilistic and thus guaranteed to notice any
errors in the result.  We discuss possibilities for improvement and possible
lower bounds in the section on future work at the end of the paper.

\subsection{Median Aggregation}\label{ss:median}

We use the common definition of the median of an even number of elements as the
mean of the two middle elements.  Thus there may not exist an element that is
equal to the median, but---assuming unique values---the number of elements
\emph{smaller} than the median is always equal to the number of \emph{larger}
elements.

If the asserted medians are available at every PE, checking a median aggregation
operation on unique values can be reduced to the sum aggregation problem by
verifying the above property.  Simply map elements smaller than their key's
median to $-1$, and larger elements to $+1$.  Then, the sum over all of these
values must be $0$ for every key.  This property can be checked
probabilistically using the sum aggregation checker of Theorem~\ref{thm:sumagg}.
Pseudocode of our algorithm is given in Algorithm~\ref{alg:median}.

Requiring that each value occur no more than once for each key is without loss
of generality because it can be enforced by an appropriate tie breaking scheme.
However, the median aggregation algorithm does not necessarily need to apply the tie
breaking scheme during its execution: we only need to know which occurrence of
the median value is the one with rank $n/2$, as determined by the chosen tie
breaking scheme.  Depending on the median aggregation algorithm used\footnote{We
  describe a communication efficient algorithm in \cite{HubSan16}, which could
  be executed in parallel for every key}, this can be much simpler than using
the tie breaking scheme in the entire algorithm.

\renewcommand{\figurename}{Algorithm}
\begin{figure}
\begin{code}
  \Funct{checkMedian}{%
    \Declare{$v$}{Element\BR{$n_i$}},
    \Declare{$o$}{Map$\langle$Key$\rightarrow$Value$\rangle$},
    \Declare{$\delta$}{$\real$}}{Boolean}\+\\
  \Declare{$s$}{Map$\langle$Key $\rightarrow \integer\rangle$ using default value 0 for unknown keys, size $k$}\\
  \Foreach{$\keyval \in v$}\+\RRem{combine mapping and local reduction}\\
  \If{\textit{val} $<o$[\textit{key}]} \Then $s$[\textit{key}] = $s$[\textit{key}]$\,- 1$\\
  \If{\textit{val} $>o$[\textit{key}]} \Then $s$[\textit{key}] = $s$[\textit{key}]$\,+ 1$\-\\

  \Return \text{checkSumAgg}($s$.to\_array$, \seq{0,\ldots,0}, \delta$)
\end{code}
\caption{\label{alg:median} Pseudocode for the median checker (unique values).
  The function \op{checkSumAgg} is defined in Algorithm~\ref{alg:sumagg}.  Here,
  $v$ is the input to the median aggregation operation, a sequence of $n$
  elements of which PE $i$ holds $n_i$; $o$ is the asserted result of size $k$;
  and $\delta$ is the maximum allowed failure rate.  For simplicity, we use
  associative arrays indexed by key for~$o$ and~$s$.}
\end{figure}
\renewcommand{\figurename}{Fig.}

\begin{thm}\label{thm:median_unique}
  Median aggregation on a sequence of $n$ elements can be checked with failure
  probability at most \mbox{$\delta>0$} in time $\Oh{\T{check-sum}(n,p,\delta)}$
  provided the asserted result is available at every PE.  For non-unique values,
  tie breaking information on the median values is required as a certificate.
\end{thm}
\begin{IEEEproof}
  By definition, an element is the median of a set of unique elements if and
  only if the number of elements smaller than it is exactly equal to the number
  of elements that is larger.  This is verified using the sum aggregation
  checker by mapping smaller values to $-1$ and larger values to $+1$.  Ties are
  broken using an appropriate tie breaking scheme and the certificate.  The
  claim then follows from Theorem~\ref{thm:sumagg}.
\end{IEEEproof}

\subsection{Zip}\label{ss:zip}

\emph{Zipping} combines two sequences $S_1=\seq{x_1,\ldots,x_n}$ and
$S_2=\seq{y_1,\ldots,y_n}$ of equal length $n$ index-wise, producing as its
result a sequence of pairs $S=\seq{(x_1,y_1), \ldots, (x_n,y_n)}$.  However,
since $S_1$ and $S_2$ need not have the same data distribution over the PEs,
this is nontrivial.  Thus, the elements of (at least) one sequence need to be
moved in the general case.  Checking \op{Zip} therefore requires verifying that
the order of the elements is unchanged in both sequences.  For this, we require
a hash function that can be evaluated in parallel on distributed data,
independent of how the input is split over the PEs.  One example would be the
inner product of the input and a sequence of $n$ random values,
$R=\seq{r_1,\ldots,r_n}$, where $r_i=h'(i)$ for some high-quality hash function
$h'$~\cite{MehSanPar}.  This way, $R$ can be computed on the fly and without
communication.

\begin{thm}\label{thm:zip}
  Checking \op{Zip(S_1, S_2)} with $\card{S_1}=\card{S_2}=n$ with false
  positive probability of at most $\delta>0$ can be achieved in time
  $\Oh{\frac{n}{pw}\log\frac1\delta + \Tbit \log \frac1\delta + \Tstart \log p}$.
\end{thm}
\begin{IEEEproof}
  A single iteration of the checker works as follows.  Apply the hash function
  to $S_1$ and the first part of the elements of $S$.  Computing this hash value
  is possible in time $\Oh{n/p + \Tbit \log H + \Tstart\log p}$ for the families
  of hash functions discussed above, where the output of the hash function is in
  $0..H-1$.  The same is applied to $S_2$ and the second part of the elements
  of~$S$.  Accept if both pairs of hashes match.  Again, the success probability
  can be boosted to $\delta$ by executing $\log_{H^{-1}}\frac1\delta$ instances of the
  checker in parallel, and the claimed running time follows for $H=2^w$.
\end{IEEEproof}

\subsection{Other Operations}

We now briefly describe several further operations to which the permutation and
sort checker can be adapted.  For the latter two, we present \emph{invasive}
checkers for the element redistribution phase.  The rest of the operation needs
to be checked with an appropriate local checker.

\subsubsection{Union}\label{ss:union}

Verifying whether a multiset $S$ is the union of two other multisets $S_1$ and
 $S_2$ of size $n_1$ and $n_2$, respectively, is equivalent to checking whether
$S$ is a permutation of the concatenation of $S_1$ and $S_2$.  We can therefore
adapt the permutation checker of Section~\ref{s:sort} to iterate over \emph{two} input
sets.
\begin{cor}\label{cor:union}
  The \op{Union(S_1, S_2)} operation can be checked in
  time~$\Oh{\T{check-perm}(\card{S_1}+\card{S_2},p,\delta)}$, where
  $\T{check-perm}(n,p,\delta)$ is the the time to check permutations of
  size~$n$ from Theorem~\ref{thm:perm}.
\end{cor}

\subsubsection{Merge}\label{ss:merge}

A \op{Merge} operation combines two sorted sequences $S_1$ and $S_2$ of length
$n_1$ and $n_2$, respectively, into a single sorted sequence $S$ of length
$n_1+n_2$.  Checking it is thus equal to verifying that $S$ is sorted and the
union of $S_1$ and $S_2$, \emph{c.f.} the previous subsection.
\begin{cor}
  Checking \op{Merge(S_1,S_2)} with probability at least $1-\delta$ is
  possible in time
  \[\Oh{\T{check-sort}(\card{S_1} + \card{S_2}, p, \delta)}.\]
\end{cor}

\subsubsection{GroupBy}\label{ss:groupby}

The \op{GroupBy} operation is conceptually similar to aggregation (see
Section~\ref{s:sumagg}), but passes all elements with the same key to the
\emph{group function} $g: [\textit{Value}] \rightarrow \textit{Value}$.  Thus,
for every key, all elements associated with it need to be sent to a single PE.
This stage of a \op{GroupBy} operation can therefore be checked using the sort
checker, where the order is induced by the hash function assigning keys to PEs.
The group function needs to be checked separately by an appropriate \emph{local
  checker}, which falls outside the scope of this paper.

\begin{cor}\label{cor:groupby}
  Checking the redistribution phase of \op{GroupBy} on a sequence of $n$
  elements with probability at least $1-\delta$ is possible in time
  $\Oh{\T{check-sort}(n, p, \delta)}$.
\end{cor}

\subsubsection{Join}\label{ss:join}

Similarly to \op{GroupBy}, we can design an invasive checker for element
redistribution in \op{Join} operations.  The two common approaches to joins are
the \emph{sort-merge join} and \emph{hash join} algorithm~\cite{DBSys}.  Note
that the sort checker can be used for both approaches, because as far as data
redistribution is concerned, a hash join is essentially a sort-merge join using
the hashes of the keys for sorting.  To further verify that the distribution of
keys to PEs is the same for both input sequences, we exchange the locally
largest (smallest) keys with the following (preceding) PE and check that those
are larger (smaller) than the local maximum (minimum).  The correctness of the
element redistribution then follows from their global sortedness.

\begin{cor}\label{cor:join}
  Checking the input redistribution phase of a hash or sort-merge \op{Join} on
  two sequences of $n_1$ and $n_2$ elements with success probability at least
  $1-\delta$ is possible in time $\Oh{\T{check-sort}(n_1+n_2, p, \delta)}$.
\end{cor}

\section{Experiments}\label{s:exp}

We developed implementations of our core checkers, sum aggregation and sorting.
These were integrated into Thrill~\cite{Thrill}, an open source data-parallel
big data processing framework\footnote{See \url{http://project-thrill.org}
  and \url{https://github.com/thrill/thrill/} for details.  Our code is available at
  \url{https://github.com/lorenzhs/thrill/tree/checkers}} using modern C++
and developed at Karlsruhe Institute of Technology.  Thrill provides a
high-performance environment that allows for easy implementation and testing of
our methods, while offering a convenient high-level interface to users.

\paragraph{Goals} The aim of our experiments is twofold.  First, to show
that our checkers achieve the predicted detection accuracy, and second, to
demonstrate their practicability by showing that very little overhead is
introduced by using a checker.

\paragraph{Manipulators}  To test the efficacy of our checkers, we implemented
\emph{manipulators} that purposefully interfere with the computation and
deliberately introduce faults.  Manipulators are a flexible way to introduce a
wide variety of classes of faults, allowing us to test our hypotheses on which
kinds of faults are hardest to detect.  It is easy to convince oneself that large-scale
manipulation of the result is much easier to detect than subtle changes.  Thus,
our manipulators focus on the latter kind of change in the data.  The specific
manipulation techniques used with an operation are described in the respective
subsections.

\paragraph{Implementation Details} As hash functions, we used CRC-32C, which
is implemented in hardware in newer x86 processors with support for SSE~4.2%
~\cite{IntelCRC}, and tabulation hashing~\cite{CarWeg81hash,PatTho12} with 256 entries per
table and four or eight tables for 32 and 64-bit values, respectively.  We
abbreviate the hash functions with ``CRC'', ``Tab'', and ``Tab64''.  Where
needed, pseudo-random numbers are obtained from an MT19937 Mersenne
Twister~\cite{MatNis98}.

\paragraph{Platform} We conducted our scaling experiments on up
to 128 nodes of bwUniCluster, each of which features two 14-core Intel Xeon
E5--2660 v4 processors and 128\,GiB of DDR4 main memory.  For our overhead and
accuracy experiments, we used an AMD Ryzen 7 1800X octacore machine with 64\,GiB
of DDR4 RAM.  The code was compiled with \texttt{g++} 7.1.0 in release mode.

\subsection{Sum Aggregation}

\begin{table}[bt]
  \centering
  \captionsetup{justification=centering}
  \caption{Configurations tested for Sum Aggregation checker.\\First set was
    used for accuracy tests, second set for scaling tests.}
\footnotesize
\begin{tabular}{lrrl}
  \toprule
  Configuration & Table size & Failure rate & Comment \\
  (\#its$\, \times d$ m$\,\log\rmin$) & (bits) & ($\delta$) \\\midrule
  $1\TI2$ m$31$ & 64 & $5\cdot10^{-1}$ & High $\rmin$ is less effective than \\
  $1\TI4$ m$31$ & 128 & $2.5\cdot10^{-1}$ & $\rcurvearrowse$ multiple iterations \\
  $4\TI2$ m$4$ & 40 & $1\cdot10^{-1}$ & Lower $\delta$ \emph{and} size than above\\
  $4\TI4$ m$3$ & 64 & $2\cdot10^{-2}$ & $\delta = 2\,\%$ for 64-bit table\\
  $4\TI4$ m$5$ & 96 & $6\cdot10^{-3}$ &\\
  $4\TI8$ m$3$ & 128 & $3.9\cdot10^{-3}$ &\\
  $4\TI8$ m$5$ & 192 & $6\cdot10^{-4}$ &\\
  $4\TI8$ m$7$ & 256 & $3.1\cdot10^{-4}$ &\\\midrule
  $5\TI16$ m$5$ & $480$ & $7.2\cdot10^{-6}$ & \\
  $6\TI32$ m$9$ & $1\,920$ & $1.3\cdot10^{-9}$ & \\
  $8\TI16$ m$15$ & $2\,048$ & $2.3\cdot10^{-10}$ & \\
  $4\TI256$ m$15$ & $16\,384$ & $2.4\cdot10^{-10}$ & \\
  $5\TI128$ m$11$ & $7\,680$ & $3.9\cdot10^{-11}$ & \\
  $8\TI256$ m$15$ & $32\,769$ & $5.8\cdot10^{-20}$ & lower local work, larger size \\
  $16\TI16$ m$15$ & $4\,096$ & $5.4\cdot10^{-20}$ & higher local work, smaller size \\
  \bottomrule
\end{tabular}\label{tbl:sumagg_configs}
\end{table}

\begin{table}[bt]
\caption{Manipulators for Sum Aggregation Checker}\centering
\footnotesize
\begin{tabular}{ll}\toprule
  Name                & Manipulation applied                                \\\midrule
  \emph{Bitflip}      & Flips a random bit in the input                     \\
  \emph{RandKey}      & randomize the key of a random element               \\
  \emph{SwitchValues} & switches the values of two random elements          \\
  \emph{IncKey}       & increments the key of a random element              \\
  \emph{IncDec$_n$}   & acts on $2n$ elements with distinct keys, \\
  ~~$\rcurvearrowse$ using $n=1$  & incrementing the keys $n$ elements and \\
  ~~$\rcurvearrowse$ and $n=2$    & decrementing that of $n$ other elements\\\bottomrule
\end{tabular}\label{tbl:sumagg_manip}
\end{table}

We implemented the sum aggregation checker of Section~\ref{s:sumagg}.  As
workload, we chose integers distributed according to a power law distribution,
with frequency $f(k; N)=1/(kH_N)$ for the element of rank $k$.  Here, $N$ is the
number of possible elements and $H_N$ is the $N$-th harmonic number.  This
distribution naturally models many workloads, \eg wordcount over natural languages.

The tested configurations of the checker---number of iterations, number of
buckets, hash function, and modulus parameter---are listed in
Table~\ref{tbl:sumagg_configs}.  The first set of configurations was used for
testing detection accuracy, the second for scaling experiments.

\paragraph{Implementation Details} \label{ss:sumagg_impl}

To minimize local work, our implementation employs \emph{bit-parallelism}
where possible, \eg, instead of computing eight four-bit hash values, we compute
one 32-bit hash value and partition it into eight groups of four bits, which we
treat as the output of the hash functions.  This is implemented in a generic
manner to satisfy any partition of a hash value into groups.  Since 64
hash bits suffice to guarantee a failure probability of nearly $10^{-20}$ (see
Table~\ref{tbl:sumagg_configs}), evaluating a single hash function suffices in
all practically relevant configurations.

To minimize the cost of adding modulo $r$, we use 64-bit values for the buckets
internally, add normally, and perform the expensive modulo step only if the
addition would overflow.  This can be detected cheaply with a jump-on-overflow
instruction.


\begin{figure}[h!]
\centering
\begin{tikzpicture}
  \begin{groupplot}[
    group style={group size=1 by 6,vertical sep=2.5mm},
    groupplot ylabel={Failure rate / (expected maximum failure rate $\delta$)},
    groupplot xlabel={Checker configuration (syntax: \#its$ \times d$ Hashfn m$\log_2 \rmin$)},
  ]
  \pgfplotsset{
    bar width=9pt,
    tick label style={rotate=90},
    xlabel={},
    xtick=\empty,
    height=10.5em,
    width=28em,
    symbolic x coords={dummy,$1\TI2$ CRC,$1\TI4$ CRC,$4\TI2$ CRC m$4$,$4\TI4$ CRC m$3$,$4\TI4$ CRC m$5$,$4\TI8$ CRC m$3$,$4\TI8$ CRC m$5$,$4\TI8$ CRC m$7$,dummy2,dummy3,$1\TI2$ Tab,$1\TI4$ Tab,$4\TI2$ Tab m$4$,$4\TI4$ Tab m$3$,$4\TI4$ Tab m$5$,$4\TI8$ Tab m$3$,$4\TI8$ Tab m$5$,$4\TI8$ Tab m$7$},
    ymax=1.1
  }
  \nextgroupplot[ylabel=\emph{Bitflip\textcolor{white}{y\kern-0.5em}}]
  \addplot[ybar,pattern=MyFill] coordinates { ($1\TI2$ CRC,0.48744) ($1\TI2$ Tab,0.49862) ($1\TI4$ CRC,0.49596) ($1\TI4$ Tab,0.4956) ($4\TI2$ CRC m$4$,0.249177) ($4\TI2$ Tab m$4$,0.325776) ($4\TI4$ CRC m$3$,0.00705696) ($4\TI4$ CRC m$5$,0.0) ($4\TI4$ Tab m$3$,0.119968) ($4\TI4$ Tab m$5$,0.3256) ($4\TI8$ CRC m$3$,0.0353938) ($4\TI8$ CRC m$5$,0.0) ($4\TI8$ CRC m$7$,0.0) ($4\TI8$ Tab m$3$,0.0740052) ($4\TI8$ Tab m$5$,0.206282) ($4\TI8$ Tab m$7$,0.0) };

  \coordinate (A) at (axis cs:dummy,1.0);
  \coordinate (O1) at (rel axis cs:0,0);
  \coordinate (O2) at (rel axis cs:1,0);
  \draw [black,sharp plot,dashed] (A -| O1) -- (A -| O2);
  \nextgroupplot[ylabel=\emph{RandKey}]
  \addplot[ybar,pattern=MyFill] coordinates { ($1\TI2$ CRC,1.00138) ($1\TI2$ Tab,0.99698) ($1\TI4$ CRC,0.99752) ($1\TI4$ Tab,1.00012) ($4\TI2$ CRC m$4$,0.636375) ($4\TI2$ Tab m$4$,0.647449) ($4\TI4$ CRC m$3$,0.238761) ($4\TI4$ CRC m$5$,0.562106) ($4\TI4$ Tab m$3$,0.235232) ($4\TI4$ Tab m$5$,0.607463) ($4\TI8$ CRC m$3$,0.119052) ($4\TI8$ CRC m$5$,0.275043) ($4\TI8$ CRC m$7$,0.0) ($4\TI8$ Tab m$3$,0.0514819) ($4\TI8$ Tab m$5$,0.275043) ($4\TI8$ Tab m$7$,0.0) };

  \coordinate (A) at (axis cs:dummy,1.0);
  \coordinate (O1) at (rel axis cs:0,0);
  \coordinate (O2) at (rel axis cs:1,0);
  \draw [black,sharp plot,dashed] (A -| O1) -- (A -| O2);
  \nextgroupplot[ylabel=\emph{SwitchValues\textcolor{white}{y}}]
  \addplot[ybar,pattern=MyFill] coordinates { ($1\TI2$ CRC,0.3357) ($1\TI2$ Tab,0.33736) ($1\TI4$ CRC,0.33452) ($1\TI4$ Tab,0.33676) ($4\TI2$ CRC m$4$,0.226823) ($4\TI2$ Tab m$4$,0.222311) ($4\TI4$ CRC m$3$,0.0911524) ($4\TI4$ CRC m$5$,0.226786) ($4\TI4$ Tab m$3$,0.107031) ($4\TI4$ Tab m$5$,0.233266) ($4\TI8$ CRC m$3$,0.0804404) ($4\TI8$ CRC m$5$,0.0859509) ($4\TI8$ CRC m$7$,0.0) ($4\TI8$ Tab m$3$,0.0514819) ($4\TI8$ Tab m$5$,0.120331) ($4\TI8$ Tab m$7$,0.0) };

  \coordinate (A) at (axis cs:dummy,1.0);
  \coordinate (O1) at (rel axis cs:0,0);
  \coordinate (O2) at (rel axis cs:1,0);
  \draw [black,sharp plot,dashed] (A -| O1) -- (A -| O2);
  \nextgroupplot[ylabel=\emph{IncKey}]
  \addplot[ybar,pattern=MyFill] coordinates { ($1\TI2$ CRC,0.4378) ($1\TI2$ Tab,0.9996) ($1\TI4$ CRC,0.5006) ($1\TI4$ Tab,1.00492) ($4\TI2$ CRC m$4$,0.329365) ($4\TI2$ Tab m$4$,0.647757) ($4\TI4$ CRC m$3$,0.0058808) ($4\TI4$ CRC m$5$,0.0032398) ($4\TI4$ Tab m$3$,0.238761) ($4\TI4$ Tab m$5$,0.690078) ($4\TI8$ CRC m$3$,0.00643523) ($4\TI8$ CRC m$5$,0.0) ($4\TI8$ CRC m$7$,0.0) ($4\TI8$ Tab m$3$,0.0965285) ($4\TI8$ Tab m$5$,0.481325) ($4\TI8$ Tab m$7$,0.0) };

  \coordinate (A) at (axis cs:dummy,1.0);
  \coordinate (O1) at (rel axis cs:0,0);
  \coordinate (O2) at (rel axis cs:1,0);
  \draw [black,sharp plot,dashed] (A -| O1) -- (A -| O2);
  \nextgroupplot[ylabel=\emph{IncDec$_1$\textcolor{white}{y\kern-0.5em}}]
  \addplot[ybar,pattern=MyFill] coordinates { ($1\TI2$ CRC,1.05462) ($1\TI2$ Tab,1.00158) ($1\TI4$ CRC,1.041) ($1\TI4$ Tab,0.99408) ($4\TI2$ CRC m$4$,0.584796) ($4\TI2$ Tab m$4$,0.648065) ($4\TI4$ CRC m$3$,0.132318) ($4\TI4$ CRC m$5$,0.3337) ($4\TI4$ Tab m$3$,0.238761) ($4\TI4$ Tab m$5$,0.597744) ($4\TI8$ CRC m$3$,0.0193057) ($4\TI8$ CRC m$5$,0.0687607) ($4\TI8$ CRC m$7$,0.0) ($4\TI8$ Tab m$3$,0.0675699) ($4\TI8$ Tab m$5$,0.343803) ($4\TI8$ Tab m$7$,0.0) };

  \coordinate (A) at (axis cs:dummy,1.0);
  \coordinate (O1) at (rel axis cs:0,0);
  \coordinate (O2) at (rel axis cs:1,0);
  \draw [black,sharp plot,dashed] (A -| O1) -- (A -| O2);
  \nextgroupplot[ylabel=\emph{IncDec$_2$\textcolor{white}{y\kern-0.5em}}, xtick=data]
  \addplot[ybar,pattern=MyFill] coordinates { ($1\TI2$ CRC,0.76072) ($1\TI2$ Tab,0.74946) ($1\TI4$ CRC,0.46572) ($1\TI4$ Tab,0.43928) ($4\TI2$ CRC m$4$,0.189908) ($4\TI2$ Tab m$4$,0.208468) ($4\TI4$ CRC m$3$,0.00764504) ($4\TI4$ CRC m$5$,0.0259184) ($4\TI4$ Tab m$3$,0.0088212) ($4\TI4$ Tab m$5$,0.0178189) ($4\TI8$ CRC m$3$,0.0) ($4\TI8$ CRC m$5$,0.0) ($4\TI8$ CRC m$7$,0.0) ($4\TI8$ Tab m$3$,0.0) ($4\TI8$ Tab m$5$,0.0) ($4\TI8$ Tab m$7$,0.0) };

  \coordinate (A) at (axis cs:dummy,1.0);
  \coordinate (O1) at (rel axis cs:0,0);
  \coordinate (O2) at (rel axis cs:1,0);
  \draw [black,sharp plot,dashed] (A -| O1) -- (A -| O2);
\end{groupplot}
\end{tikzpicture}
\caption{Accuracy of the Sum Aggregation checker for different manipulators.
  The $50\,000$ input elements follow a power law distribution with $10^6$
  possible values, executed on 4~PEs and measured over $100\,000$ iterations.}
\label{fig:reduce_accuracy}
\end{figure}
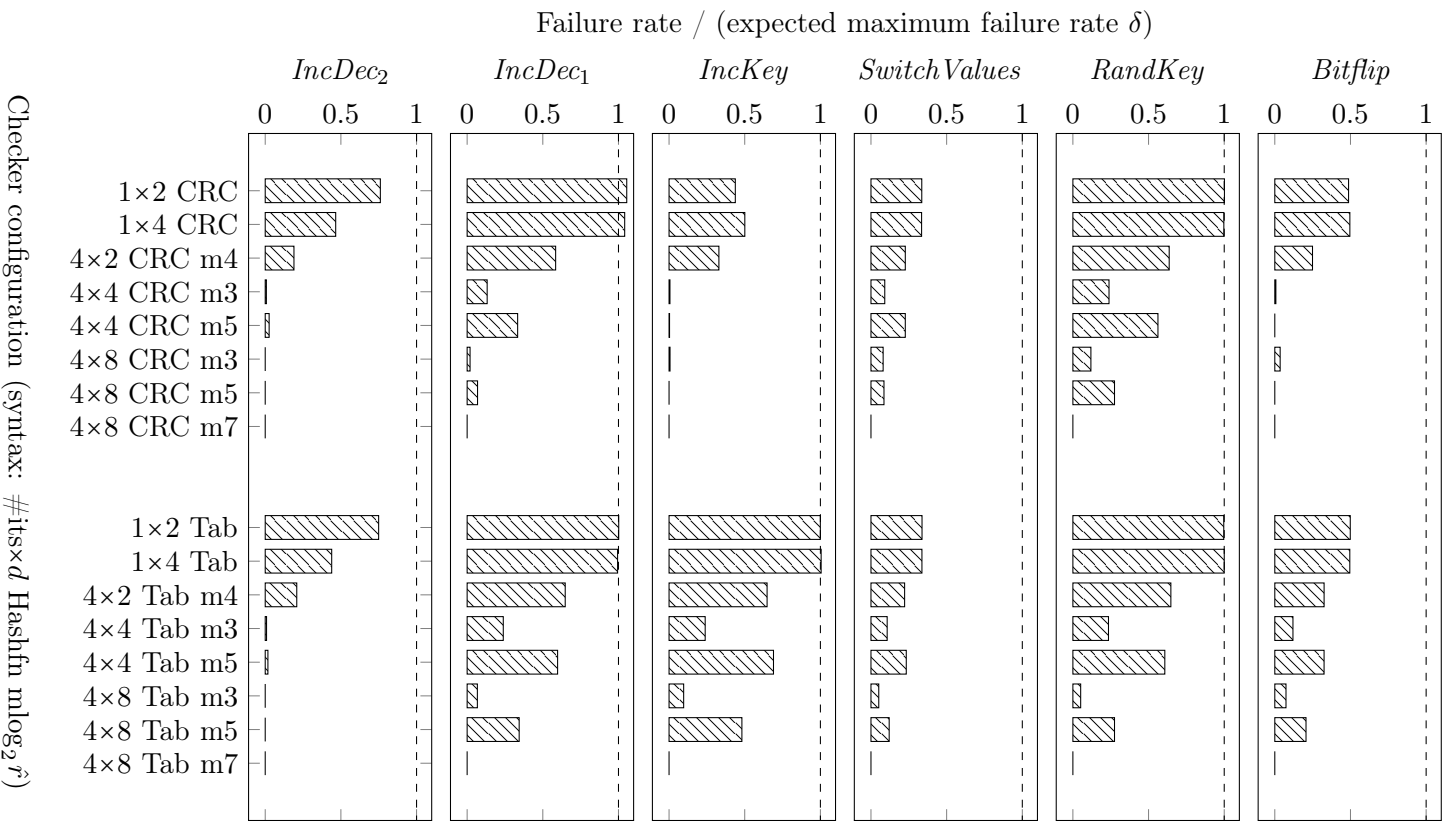


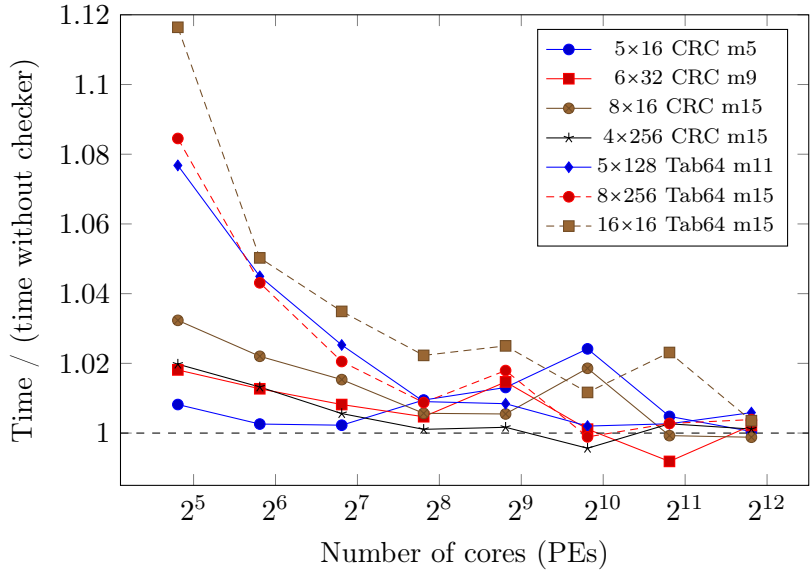
\begin{figure}[bt]
\centering
\begin{tikzpicture}
  \begin{semilogxaxis}[
    width=0.65\textwidth,
    height=0.475\textwidth,
    log basis x=2,
    ymax=1.12,
    ymin=0.985,
    legend pos=north east,
    legend style={font=\scriptsize},
    xlabel={Number of cores (PEs)},
    ylabel={Time / (time without checker)}
  ]
  \addplot coordinates { (28,1.00819) (56,1.00259) (112,1.00226) (224,1.00948) (448,1.01308) (896,1.02416) (1792,1.00479) (3584,1.00046) };
  \addlegendentry{$5\TI16$ CRC m$5$};
  \addplot coordinates { (28,1.01811) (56,1.01271) (112,1.00819) (224,1.00469) (448,1.01467) (896,1.00113) (1792,0.991872) (3584,1.00218) };
  \addlegendentry{$6\TI32$ CRC m$9$};
  \addplot coordinates { (28,1.03234) (56,1.02203) (112,1.01534) (224,1.00567) (448,1.00548) (896,1.01858) (1792,0.999267) (3584,0.998809) };
  \addlegendentry{$8\TI16$ CRC m$15$};
  \addplot coordinates { (28,1.01977) (56,1.01322) (112,1.0056) (224,1.00107) (448,1.00166) (896,0.995647) (1792,1.00266) (3584,1.00116) };
  \addlegendentry{$4\TI256$ CRC m$15$};
  \addplot coordinates { (28,1.07679) (56,1.04496) (112,1.02527) (224,1.00903) (448,1.00843) (896,1.00201) (1792,1.00267) (3584,1.00587) };
  \addlegendentry{$5\TI128$ Tab64 m$11$};
  \addplot coordinates { (28,1.08452) (56,1.04311) (112,1.02053) (224,1.00875) (448,1.01794) (896,0.998909) (1792,1.00276) (3584,1.00387) };
  \addlegendentry{$8\TI256$ Tab64 m$15$};
  \addplot coordinates { (28,1.11642) (56,1.05025) (112,1.03491) (224,1.02229) (448,1.02501) (896,1.01166) (1792,1.02312) (3584,1.00356) };
  \addlegendentry{$16\TI16$ Tab64 m$15$};

  \coordinate (A) at (axis cs:9999,1.0);
  \coordinate (O1) at (rel axis cs:0,0);
  \coordinate (O2) at (rel axis cs:1,0);
  \draw [black,sharp plot,dashed] (A -| O1) -- (A -| O2);
 \end{semilogxaxis}
\end{tikzpicture}
\captionsetup{justification=centering}
\caption{Weak scaling experiment for Sum Aggregation checker with
  $125\,000$ items per PE,\\following a Zipf (power law) distribution. Average
  over 1000 runs.}
\label{fig:reduce_scaling}
\end{figure}

\paragraph{Detection Accuracy} The manipulators we used are listed in
Table~\ref{tbl:sumagg_manip}.  Fig.~\ref{fig:reduce_accuracy} demonstrates that
our checkers achieve the theoretically predicted performance in practice, even
when using hash functions with weaker guarantees.  We can see that
Lemma~\ref{lem:sumagg_prob} generally overestimates the failure probability introduced by the
modulus.  Note that the significance of the results for configurations with low
$\delta$ is limited, as the expected number of failures in
$100\,000$ runs for \eg $4\TI8$ Tab m$7$ is only~$31.1$.


We note that CRC-32C appears to work very well for subtle manipulations, which
is likely because it was designed so that small changes in the input cause many
output bits to change.  This makes it likely that a change in an element's key
causes it to be sent to a different bucket in at least one iteration.  However,
the least significant bits appear to change in similar ways for different
inputs, as indicated by the \emph{IncDec$_1$} measurements, causing an elevated
failure rate in this setting.  Tabulation hashing performs quite uniformly well
across the board, which is not entirely unexpected given its high degree of
independence.

\begin{table}\centering
  \captionsetup{justification=centering}
  \caption{Overhead of sum aggregation checker: checker local input\\processing
    time for $10^6$ pairs of 64-bit integers, $10\,000$ runs}
  \label{tbl:wc_overhead}
  \footnotesize
\begin{tabular}{lrr}\toprule
  Configuration (see Table~\ref{tbl:sumagg_configs}) & Time per element [ns] & Approx. \#cycles \\\midrule
    $5\TI16$ CRC m$5$ &  4.5 & 16 \\
    $6\TI32$ CRC m$9$ &  4.6 & 17 \\
   $8\TI16$ CRC m$15$ &  5.1 & 18 \\
  $4\TI256$ CRC m$15$ &  3.8 & 14 \\
$5\TI128$ Tab64 m$11$ &  4.7 & 17 \\
$8\TI256$ Tab64 m$15$ &  7.3 & 26 \\
$16\TI16$ Tab64 m$15$ & 10.0 & 36 \\
  \bottomrule
\end{tabular}
\end{table}
\def\wcseq{88}

\paragraph{Overhead}
We measured the sequential overhead of the sum aggregation checker on
the aforementioned AMD machine, as single-node performance on bwUniCluster proved
to be too inconsistent to yield useful results (this is likely due to thermal limitations).  The results are shown in
Table~\ref{tbl:wc_overhead}.  The configurations used provide high confidence in
the correctness of the result and cover a wide range of practically relevant
parameter choices.  We can see that $\delta < 10^{-10}$ can be achieved with less than
5\,ns per element on this 3.6\,GHz machine. A configuration with very
small table size (512 Bytes) and thus very little communication, with
$\delta \approx 5\cdot 10^{-20}$, can be
realized with as little as 10\,ns per element, approximately 36 cycles.  For
comparison, the main reduce operation takes approximately \wcseq\,ns per element
using a single core of the same machine.  This surprisingly low overhead was
achieved by carefully engineering our implementation as described in
Section~\ref{ss:sumagg_impl}.  A naive implementation would likely cause at
least one order of magnitude more overhead.

\def\wcoverhead{1.1}

\def\wcoverheadp{2.4}

\paragraph{Scaling Behavior} We performed weak scaling experiments, the
results of which are shown in Fig.~\ref{fig:reduce_scaling}.  These measurements
suffer from a noticeable amount of noise because reduction and checker are
interleaved---elements are forwarded to the checker as they are passed to the
reduction.  This is necessitated by the design of Thrill, as in related Big Data
frameworks.  As a result, we measure the entire reduce-check pipeline, the
running time of which is influenced by multiple sources of variability,
including the network, causing the aforementioned noise.
We note that the results for a single node remain consistent with the sequential
overhead measurements of the preceding paragraph.

It is clearly visible that the overhead introduced by the checkers
is within the fluctuations introduced by the network and other sources of noise.
Furthermore, starting with four nodes, data exchange for the reduction
dominates overall running time, which becomes nearly independent from the
checker's configuration and thus accuracy, although some impact of the number of
iterations on running time remains visible.  Nonetheless, the average overhead
over all configurations is a mere $\wcoverhead\,\%$ when using more
than a single node.  Observe that even the slowest and most accurate configuration,
``$16\TI16$ Tab64 m$15$'' with $\delta < 6\cdot10^{-20}$, thus providing
near-certainty in the correctness of the result, adds only $\wcoverheadp\,\%$ to
the average running time on two or more nodes.

\subsection{Permutation and Sorting}\label{s:exp:sort}

\begin{table}[t]
\caption{Manipulators for Sort/Permutation Checker}\centering
\footnotesize
\begin{tabular}{ll}\toprule
  Name             & Manipulation applied                        \\\midrule
  \emph{Bitflip}   & Flips a random bit in the input             \\
  \emph{Increment} & increment some element's value              \\
  \emph{Randomize} & set some element to a random value          \\
  \emph{Reset}     & reset some element to the default value (0) \\
  \emph{SetEqual}  & set some element equal to a different one   \\\bottomrule
\end{tabular}\label{tbl:perm_manip}
\end{table}


\def\sortoverhead{3.5}



In this subsection, we evaluate the permutation and sorting checker of
Section~\ref{s:sort} with a workload of $10^6$ integers chosen uniformly at
random from the interval $0..10^8-1$.  We implemented a single iteration of the
hash-based permutation and sorting checker of Lemma~\ref{lem:perm_hash}, and
truncate the output of the hash function to $H$ bits for different values of
$H$.  Manipulations are applied \emph{before} sorting in order to test the
permutation checker and not the trivial sortedness check.

\paragraph{Detection Accuracy} We measured detection accuracy of the sort
checker using CRC-32C and tabulation hashing, two fast real-world hash functions
with limited randomness.  Our experiments in Appendix~\ref{app:sortacc} show
that tabulation hashing provides sufficient randomness to achieve the
theoretically predicted detection accuracy on all tested manipulators.  We
further observe that CRC-32C does \emph{not} seem to provide sufficient
randomness for all of the manipulators listed in Table~\ref{tbl:perm_manip}.  In
particular, we observed a significantly increased failure rate using the
\emph{Increment} manipulator where a single element of the input was incremented
by 1.  No significant deviations from expected performance were observed for the
other manipulators.

\paragraph{Running Time} On the aforementioned AMD machine, the overhead for
local processing of input and output of the sorting operation was 2.0\,ns per
element for CRC32, and 2.8\,ns when using 32-bit tabulation hashing.  This
corresponds to roughly $\sortoverhead\,\%$ of total running time when
considering 100\,000 elements.  As the time spent on the hash function does not
depend on how many of its output bits are used, the configuration did not have
measurable impact on the running time.

We do not present scaling experiments for the permutation checker, as the only
communication is a global reduction on a single integer value and one message
sent and received per PE, both also containing exactly one integer value.

\section{Conclusions}\label{s:conclusions}

We have shown that probabilistic checking of many distributed big data
operations is possible in a communication efficient manner.  Our experiments
show excellent scaling and that the running time overhead of the checkers in a distributed setting is
below $5\,\%$ for sum aggregation and sorting, even when near-certainty in the
correctness of the result is required.  Accuracy guarantees predicted by
theoretical analysis are achieved in practice even when using hash functions
with limited randomness, and memory overhead is negligible.  The basic methods
used in our checkers are extremely simple, making manual verification of the
correctness of the checker feasible.  The existence of such checkers could speed
up the development cycles of operations in big data processing frameworks by
providing correctness checks and allowing for graceful degradation at execution
time by falling back to a simpler but slower method should a computation fail.

\subsubsection*{Future Work} Lower bounds on communication volume and latency
for probabilistic distributed checkers would offer some insight into how far
from an optimal solution our checkers are.  Furthermore, it would be
interesting to see whether more operations can be checked without the need for a
certificate or requiring the purported result to be available at each PE.  For
example, could a probabilistic minimum aggregation checker with sublinear
communication exist?

It would also be interesting to know whether the sum aggregation checker can be
adapted for other data types such as floating point numbers without suffering
from numerical instability issues such as catastrophic cancellation.


\subsubsection*{Acknowledgments} The authors acknowledge support by the state
of Baden-Württemberg through bwHPC.  We would like to thank Timo Bingmann for
support with Thrill.


\bibliographystyle{IEEEtranN}
\bibliography{diss}

\clearpage
\begin{appendix}
\section{Permutation Checker Accuracy}\label{app:sortacc}

\begin{figure}[h]
\centering
\begin{tikzpicture}
  \begin{groupplot}[
    group style={group size=1 by 5,vertical sep=2.5mm},
    groupplot ylabel={Failure rate / (expected maximum failure rate $\delta$)},
    groupplot xlabel={Checker configuration (syntax: Hashfn $\log H$, i.e. \#bits)},
  ]
  \pgfplotsset{
    height=10em,
    width=26em,
    xlabel={},
    symbolic x coords={dummy,Tab $1$,Tab $2$,Tab $3$,Tab $4$,Tab $6$,Tab $8$,Tab $12$,dummy2,dummy3,CRC $1$,CRC $2$,CRC $3$,CRC $4$,CRC $6$,CRC $8$, CRC $12$},
    xtick=\empty,
    tick label style={rotate=90},
    ymin=0,
    bar width=8pt
  }
  \nextgroupplot[ylabel=\emph{Bitflip}]
  \addplot[ybar,pattern=MyFill] coordinates { (CRC $1$,1.00152) (CRC $12$,0.98304) (CRC $2$,0.9992) (CRC $3$,1.00792) (CRC $4$,0.9864) (CRC $6$,1.03168) (CRC $8$,1.01376) (Tab $1$,0.99982) (Tab $12$,1.26976) (Tab $2$,0.99012) (Tab $3$,0.98936) (Tab $4$,0.98192) (Tab $6$,0.9856) (Tab $8$,0.89344) };

  \coordinate (A) at (axis cs:dummy,1.0);
  \coordinate (O1) at (rel axis cs:0,0);
  \coordinate (O2) at (rel axis cs:1,0);

  \draw [black,sharp plot,dashed] (A -| O1) -- (A -| O2);
  \nextgroupplot[ylabel=\emph{Increment\textcolor{white}{p}\kern-0.5em},ytick={0,3,6}]
  \addplot[ybar,pattern=MyFill] coordinates { (CRC $1$,1.02926) (CRC $12$,6.9632) (CRC $2$,1.09104) (CRC $3$,1.42192) (CRC $4$,2.43824) (CRC $6$,2.39552) (CRC $8$,3.07968) (Tab $1$,0.9954) (Tab $12$,0.69632) (Tab $2$,0.9896) (Tab $3$,1.00128) (Tab $4$,0.99424) (Tab $6$,1.00544) (Tab $8$,1.02144) };

  \coordinate (A) at (axis cs:dummy,1.0);
  \coordinate (O1) at (rel axis cs:0,0);
  \coordinate (O2) at (rel axis cs:1,0);

  \draw [black,sharp plot,dashed] (A -| O1) -- (A -| O2);
  \nextgroupplot[ylabel=\emph{Randomize\textcolor{white}{p}\kern-0.5em}]
  \addplot[ybar,pattern=MyFill] coordinates { (CRC $1$,0.99922) (CRC $12$,0.98304) (CRC $2$,0.9952) (CRC $3$,0.9996) (CRC $4$,0.98896) (CRC $6$,1.00096) (CRC $8$,1.1008) (Tab $1$,0.99994) (Tab $12$,0.77824) (Tab $2$,1.00496) (Tab $3$,1.00792) (Tab $4$,1.01344) (Tab $6$,0.97792) (Tab $8$,0.97536) };

  \coordinate (A) at (axis cs:dummy,1.0);
  \coordinate (O1) at (rel axis cs:0,0);
  \coordinate (O2) at (rel axis cs:1,0);

  \draw [black,sharp plot,dashed] (A -| O1) -- (A -| O2);
  \nextgroupplot[ylabel=\emph{Reset\textcolor{white}{p}\kern-0.5em}]
  \addplot[ybar,pattern=MyFill] coordinates { (CRC $1$,0.99358) (CRC $12$,0.86016) (CRC $2$,1.0) (CRC $3$,1.0144) (CRC $4$,1.00176) (CRC $6$,1.056) (CRC $8$,1.03936) (Tab $1$,0.99898) (Tab $12$,0.90112) (Tab $2$,1.00732) (Tab $3$,1.0) (Tab $4$,0.98992) (Tab $6$,0.99712) (Tab $8$,1.088) };

  \coordinate (A) at (axis cs:dummy,1.0);
  \coordinate (O1) at (rel axis cs:0,0);
  \coordinate (O2) at (rel axis cs:1,0);

  \draw [black,sharp plot,dashed] (A -| O1) -- (A -| O2);
  \nextgroupplot[ylabel=\emph{SetEqual},xtick=data]
  \addplot[ybar,pattern=MyFill] coordinates { (CRC $1$,1.00292) (CRC $12$,1.14688) (CRC $2$,0.99684) (CRC $3$,1.0068) (CRC $4$,1.01168) (CRC $6$,0.97664) (CRC $8$,0.96768) (Tab $1$,0.9995) (Tab $12$,0.73728) (Tab $2$,0.99644) (Tab $3$,1.0052) (Tab $4$,0.98384) (Tab $6$,0.94528) (Tab $8$,1.0624) };

  \coordinate (A) at (axis cs:dummy,1.0);
  \coordinate (O1) at (rel axis cs:0,0);
  \coordinate (O2) at (rel axis cs:1,0);

  \draw [black,sharp plot,dashed] (A -| O1) -- (A -| O2);
\end{groupplot}
\end{tikzpicture}
\caption{Accuracy of the \op{Permutation}/\op{Sort} checker for different
  manipulators.  Uniformly distributed input with $10^8$ possible values, $10^6$
  input elements, 4~PEs, 100\,000 iterations for each manipulator.}
\label{fig:sort_accuracy}
\end{figure}
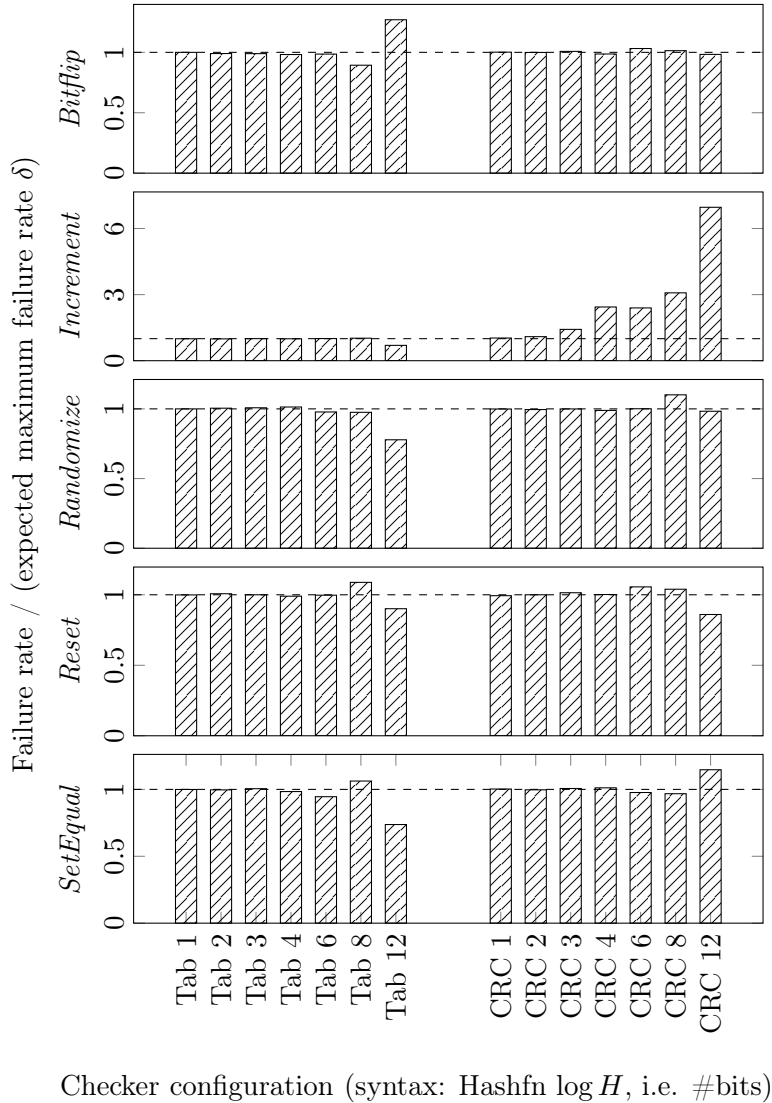

In Figure~\ref{fig:sort_accuracy} we show the results of accuracy experiments
for the permutation checker as described in Section~\ref{s:exp:sort}.  This
verifies whether the hash functions used provide sufficient randomness to detect
manipulations.  As discussed in Section~\ref{s:exp:sort}, we can see that
CRC-32C does not provide sufficient randomness to detect off-by-one errors in
element values as simulated using the \emph{Increment} manipulator (see
Table~\ref{tbl:perm_manip}).

\bigskip

\end{appendix}

\end{document}